\documentclass[12pt]{article}

\usepackage{amsmath,amssymb,bm,mathrsfs}
\usepackage{hyperref}
\usepackage{graphicx}

\newcommand{\beq}{\begin{equation}}
\newcommand{\eeq}{\end{equation}}
\newcommand{\beqn}{\begin{eqnarray}}
\newcommand{\eeqn}{\end{eqnarray}}

\newcommand{\beas}{\begin{eqnarray*}}
\newcommand{\eeas}{\end{eqnarray*}}

\newcommand{\bquo}{\begin{quote}}
\newcommand{\enqu}{\end{quote}}

\newcommand{\tN}{\tilde N}

\def\d{\partial}

\def\2{{1\over 2}}
\def\ntwo{${\mathcal N}=2\;$}

\def\ntwot{${\mathcal N}=(2,2)\;$}

\def\ba{\beq\new\begin{array}{c}}
\def\ea{\end{array}\eeq}

\newcommand{\pt}{\partial}

\begin{document}

\begin{flushright}
{FTPI-MINN-15/04, UMN-TH-3418/15 }
\end{flushright}

\vspace{3mm}

\begin{center}
{ \bf \Large Critical  String from Non-Abelian Vortex in\\[2mm]
 Four Dimensions}

\vspace{6mm}

M. Shifman$^a$ and A. Yung$^{a,b}$

\vspace{2mm}

{\em 
$^a${William I. Fine Theoretical Physics Institute, University of Minnesota,
Minneapolis, MN 55455, USA}

$^b${Petersburg Nuclear Physics Institute, Gatchina, St. Petersburg
188300, Russia\\
St. Petersburg State University,
 St. Petersburg 198504, Russia}
}

\vspace{10mm}

{\bf \large Abstract}

\end{center}

In a class of non-Abelian solitonic vortex strings  supported in  certain
\ntwo super-Yang-Mills theories we search for the vortex which can behave as a critical fundamental string.
We use the Polchinski-Strominger criterion of the ultraviolet completeness.
We identify an appropriate four-dimensional bulk theory:
it has the $U(2)$ gauge group, the Fayet-Iliopoulos term and four flavor hypermultiplets.
It supports semilocal vortices with the world-sheet theory for orientational (size) moduli
described by the weighted $CP(2,2)$ model. The latter is superconformal. Its  target space is six-dimensional. 
The overall Virasoro central charge is critical. We show that the world-sheet theory
on the vortex supported in this bulk model is the {\em bona fide} critical string.

\newpage

{\em Introduction.}---Since 2003 a large variety of non-Abelian solitonic vortices supported in certain four-dimensional
super-Yang-Mills theories were discovered \cite{1}.  Such vortices contain extra non-Abelian moduli (the so-called orientational moduli), in addition to the conventional translational moduli. The low-energy theory for the orientational moduli fields on the vortex world sheet is usually a nonlinear sigma model, typically $CP(N-1)$, with different degrees of supersymmetry. The primary purpose of the
 non-Abelian vortex explorations is modeling confinement and related phenomena in QCD-like theories.

These vortex strings are not similar to the critical strings of the fundamental string theory, and cannot be treated as such. The most clear-cut distinction is the fact that the world sheet theory is not conformal. In the terminology of Ref.  \cite{PolchStrom} such world sheet theories are not ultraviolet (UV) complete:  higher derivative terms
are needed to make them consistent in the UV.

In this Letter we report the following finding. A semilocal ${\mathcal N} = (2,2)$ vortex with the orientational moduli 
described by the weighted $CP(2,2)$ model {\em is} UV complete, and  reduces 
to a critical string in four dimensions. 

This vortex is supported in four-dimensional ${\mathcal N} = 2$ super-Yang-Mills
with the $U(2)$ gauge group, the Fayet-Iliopoulos term, and four flavor hypermultiplets. 
The target space metric in the world sheet theory has a block form: four-by-four block in the upper left corner (corresponding to flat metric for translational moduli) and six-by-six block in the lower right corner (corresponding to 
a Calabi-Yau metric with the vanishing Ricci tensor for orientational moduli). This metric can be read off from Eq. (\ref{wcp}). The world-sheet theory is conformally invariant. 

{\em General considerations.}---It is known that the hadron spectrum is well described by  linear Regge trajectories. In the early days of string theory this fact motivated people to consider dual resonance models as
 a theory of hadrons. It is believed that confinement in QCD is due formation of confining strings. 
 In all known examples in which the confining strings are formed in a controllable way,
say, the Abrikosov-Nielsen-Olesen (ANO) string \cite{ANO} in the weakly coupled Abelian-Higgs model or 
the Seiberg-Witten strings  in slightly deformed \ntwo super-Yang-Mills theory \cite{SW1}, the
 Regge trajectories will show linear behavior only at asymptotically large spins \cite{Yrev00,Shifman2005}.

Indeed, consider an open string rotating with the spin $J$. In the (semi)clas\-sical approximation its
length is determined by the relation $L^2 \sim J/{T}$
where $T$ is the string tension. Its transverse size is given  by the inverse mass $m$ of the bulk fields
forming the string, say, for the ANO string the masses of  the gauge and Higgs fields  
\footnote{For BPS-saturated strings these masses are equal.}. At weak coupling these masses 
typically scale as $m\sim g\sqrt{T}$
where $g$ is a small gauge coupling constant.

Clearly the string excitation spectrum can form linear Regge trajectories only if the string length is much larger than its 
transverse size. This gives the condition
\beq
mL\gg 1\,,\,\,\mbox{or}\,\,J\gg 1/g^2\,.
\label{largeL}
\eeq
 At weak coupling $g$ is small so spin $J$ should be large.

At $J\sim 1$ the condition (\ref{largeL}) is not met at weak coupling:  the string is not developed.
Rather, we deal with a sausage-like field configuration.
The quark-antiquark mesons formed are closer to spherical symmetry. No  linear Regge trajectory
apply at $J\sim 1$.

Empirically in the real world QCD we have practically linear Regge trajectories at $J\sim 1$ .
 Can we find {\em any} example
of a four-dimensional bulk theory where confining string remains thin at $J\sim 1$?
If so,  the string must satisfy the condition
\beq
T\ll m^2\,,
\label{thinstring}
\eeq
to be referred to as the thin string condition. This condition cannot be met at weak coupling.

{\em Strong coupling.}---We have to find an appropriate strongly coupled  four-dimensional bulk theory.
We will find super-Yang-Mills theory which supports vortices similar to critical strings 
(e.g. with  conformal world sheet theory) and then  formulate a necessary conditions
for the existence of the thin string regime. 

{\em Thin string regime.}---In the effective two-dimensional theory on the string world sheet the
problem can be understood as follows.
For the ANO string the effective theory on the string world sheet 
is given by the Nambu-Goto action {\em plus} higher derivative corrections. 
Higher derivative terms are needed to make the world sheet theory UV complete  \cite{PolchStrom}. 
This requirement can
be used to constrain higher derivative terms order by order in the derivative expansion, 
see e.g. \cite{AharonyKomar} and references therein. 

Higher derivative corrections run in powers of the ratio ${\pt^2}/{m^2}$
where the mass of the bulk fields $m$ is given by $m\sim g\sqrt{T}$ and typical energy in the numerator 
at $J\sim1$ is 
determined by the string tension. Thus, higher derivative corrections materialize as powers of  
$T/m^2$. Obviously they all blow up at weak coupling -- the string surface become
``crumpled'' \cite{Polyak86}. This is the world-sheet implementation of the bulk picture of a
short and thick ``string."

We want to find a regime in which the string remains thin, see (\ref{thinstring}).
This means that the higher derivative corrections should be parametrically small.
In other words,  the low-energy world-sheet theory  \footnote{By low energy-theory we mean
a theory with no more than two derivatives in the Polyakov formulation \cite{P81}, see Eq. (\ref{wcp}).}
should be UV complete. This leads us to the  following necessary conditions to have such a regime:

\vspace{1mm}

(i)  The low-energy world-sheet theory 
 on the string must be conformally invariant;

(ii) It must have the critical value of the Virasoro central charge.

\vspace{1mm}

  These are the famous conditions satisfied by the fundamental
string. In particular, the bosonic fundamental string becomes critical in $D=26$, while the superstring becomes
critical in $D=10$. The low energy world sheet theory for the  ANO string  is not critical in four dimensions.
We will show below that the above conditions are met in a class of the non-Abelian vortices \cite{HT1,ABEKY,SYmon,HT2}. In particular, the solitonic vortex in question must have six orientational moduli,
which, together with four translational moduli, will form a ten-dimensional space.

{\em Non-Abelian vortices.}---Non-Abelian vortices are supported in a large class of supersymmetric and non-supersymmetric gauge theories.
We will focus on the bulk four-dimensional theories in which the non-Abelian vortices were first found:
 \ntwo supersymmetric QCD with the $U(N)$ gauge group, $N_f$ quark flavor multiplets ($N_f\ge N$) and the
Fayet-Iliopoulos (FI) parameter $\xi$ of the U(1) factor of the gauge group.
In this theory the vortices under consideration are BPS-saturated  and preserve half of the bulk supersymmetry.
Thus, they possess \ntwot supersymmetry on the world sheet.  The string tension is determined exactly  by
 \beq
T_P=2\pi\xi.
\label{ten}
\eeq
These strings are formed due to the (s)quark condensation; therefore, they confine monopoles. More precisely,
in the $U(N)$ gauge theories the confined monopoles are implemented as junctions of two vortices of   different kinds \cite{T,SYmon,HT2}. 

Dynamics of the translational modes in the Polyakov formulation \cite{P81} can be described by the action
\beq
S_{\rm tr.} = \frac{T}{2}\,\int d^2 \sigma \sqrt{h}\, h^{\alpha\beta}\d_{\alpha}x^{\mu}\,\d_{\beta}x_{\mu},
\label{trans}
\eeq
where $\sigma^{\alpha}$ ($\alpha=1,2$) are the world-sheet coordinates, $x^{\mu}$ ($\mu=1,...,4$) describes the string
world sheet and $h=det(h_{\alpha\beta})$ where $h_{\alpha\beta}$ is the world-sheet metric which is understood as a
independent variable \footnote{Effective world-sheet theories for both translational and orientational moduli
 are derived in the quasiclassical approximation. In this approximation two alternative string theory 
formulations -- in terms
of the induced metric and in terms of the metric $h_{\alpha\beta}$ as an independent variable -- 
are equivalent.}.

If $N_f=N$  the dynamics of the orientational zero modes of the vortex, which become orientational moduli fields 
 on the world sheet, is described by two-dimensional
\ntwot-supersymmetric $CP(N-1)$ model.
If one adds extra quark flavors, non-Abelian vortices become semilocal.
They acquire size moduli \cite{AchVas}.  

Non-Abelian semilocal vortices in \ntwo SQCD with $N_f>N$ were studied in
\cite{HT1,HT2,SYsem,Jsem,SYV}. 
The world-sheet theory for the orientational
moduli of the semilocal vortex is given\footnote{Both the orientational and the size moduli
have logarithmically divergent norms, see e.g.  \cite{SYsem}. After an appropriate infrared regularization, logarithmically divergent norms  can be absorbed into the definition of relevant two-dimensional fields  \cite{SYsem}.
In fact, the world-sheet theory on the semilocal non-Abelian string is 
not exactly the weighted $CP(N,\tN)$ model \cite{SYV}, there are minor differences unimportant for our purposes. The actual theory is called the $zn$ model. We can ignore 
the above differences.}  by the weighted $CP(N,\tN)$ sigma model
where $\tN=(N_f-N)$. Its gauged formulation is as follows \cite{W93}. One introduces
two types of complex fields, with the $U(1)$ charges $\pm 1$:
 $n^P$ ($P=1, ..., N$) and $\rho^K$ ($K=N+1, ..., N_f$). The orientational moduli are described by 
 the $N$-plets $n^P$
 while the size moduli are parametrized by the $\tN$-plet
$\rho^K$. 

The effective two-dimensional theory on the world sheet has the action
\beqn
S_{\rm or.} &=& \int d^2 \sigma \sqrt{h} \left\{ h^{\alpha\beta}\left(
 \tilde{\nabla}_{\alpha}\bar{n}_P\,\nabla_{\beta} \,n^{P} 
 +\nabla_{\alpha}\bar{\rho}_K\,\tilde{\nabla}_{\beta} \,\rho^K\right)
 \right.
\nonumber\\[3mm]
&+&\left.
 \frac{e^2}{2} \left(|n^{P}|^2-|\rho^K|^2 -2\beta\right)^2
\right\}+\mbox{fermions}\,,
\label{wcp}
\eeqn
where $P=1,..., N$ and $ K=N+1,..., N_f$.
The fields $n^{P}$ and $\rho^K$ have
charges  +1 and $-1$ with respect to the auxiliary U(1) gauge field;
hence, 
$$ \nabla_{\alpha}=\d_{\alpha}-iA_{\alpha}\,, \qquad \tilde{\nabla}_{\alpha}=\d_{\alpha}+iA_{\alpha}\,.$$
The limit $e^2\to\infty$ is implied.

The coupling constant $\beta$ in (\ref{wcp}) is related to the bulk coupling via 
\beq
\beta= {2\pi}/{g^2}\,.
\label{betag}
\eeq
Note that the first  (and the only) coefficient of the $\beta$ function $b=N-\tN$ is the same for the bulk and 
world-sheet theories.

The bosonic part of the total string action for the non-Abelian vortex under consideration is 
the sum of (\ref{trans}) and (\ref{wcp}),
\beq
S= S_{\rm tr.} + S_{\rm or.}
\label{stringaction}
\eeq
The target space of (\ref{wcp}) has dimension $2(N+\tN ) -2 = 2({N_f} -1)$. If $N=\tN=2$ the target space is six-dimen\-sional.

{\em From vortices to thin non-Abelian strings.}---We can ask whether the above vortex can satisfy
 two necessary conditions
 to become the thin string. In other words, can the string theory (\ref{stringaction}) be UV complete
in the same way as the fundamental string theory?

In the conformal gauge the translational part of the action is a free theory and therefore conformal, while
the orientational part's  $\beta$ function is proportional to $b=N-\tN$. Thus, the condition of the conformal
invariance $b=0$ implies
\beq
N=\tN,\,\, \mbox{or} \,\, N_f=2N.
\label{confinv}
\eeq

The total Virasoro central charge for the string theory (\ref{stringaction}) is given by
\beq
c_{\rm tot}= \frac32\left( D +\frac23\,c_{\rm wcp} -10\right),
\label{ccharge}
\eeq
where the first term is the contribution of the translational sector with $D=4$, $c_{\rm wcp}$ is
the Virasoro central charge of the weighted $CP(N,\tN)$  model (\ref{wcp}) in the conformal gauge, and 
the last term is the ghost contribution for the superstring. 
Our world-sheet theory has \ntwot world sheet supersymmetry which is needed to
have space-time supersymmetry in the bulk theory
\cite{Gepner,BDFM}.

To calculate the central charge of the weighted $CP(N,\tN)$  model (\ref{wcp}) we use the formula \cite{VafaWarner}
\beq
c= 3\sum_i (1-2q_i)
\label{cq}
\eeq
which relates the central charge of \ntwot sigma model to the $R$ charges.
Here the sum runs over the chiral multiplets and $q_i$ is the  $R$ charge of the given multiplet. In our theory
(\ref{wcp}) the $R$ charges of the fields $n^P$ and $\rho^K$ vanish. Thus, Eq. (\ref{cq}) just counts
the number of degrees of freedom, namely
\beq
c_{\rm wcp}= 3(N+\tN-1).
\label{ccWCP}
\eeq
Now the condition of criticality takes the form
\beqn
c_{\rm tot}&=& \frac32 \left[ D + 2(2N-1) -10\right] =0,\\[1mm]
N&=&(12-D)/4\,.
\label{critcond}
\eeqn
where we used (\ref{confinv}). For $D=4$ this condition has a solution
\beq
N=\tN=2, \qquad N_f=4.
\label{critstr}
\eeq
For these values of $N$ and $\tN$
the target space of the  weighted $CP(N,\tN)$ model (\ref{wcp}) is a noncompact 
Calabi-Yau manifold studied in  \cite{W93}.

Thus, the non-Abelian string can potentially become thin in a particular four-dimensional theory:
\ntwo supersymmetric QCD with the $U(2)$ gauge group and $N_f=4$ flavors of quarks.
Given the necessary conditions are met we conjecture that  the thin-string condition 
(\ref{thinstring}) is actually satisfied
in this theory in the strong coupling limit. 

In fact, the bulk theory at zero $\xi$ possess a strong-weak duality
$\tau \to -1/\tau$, where $\tau = i\,8\pi/g^2 + \theta/\pi$ and $\theta$ is the $\theta$-angle 
 \cite{ArgPlessShapiro,APS}. Therefore,
even at non-zero $\xi$ the region of $g^2\gg 1$ can be described in terms of the weakly coupled U$(N)$
dual gauge theory \footnote{Although the gauge coupling constant
does not run in the bulk theory  at  $N_f=2N$ the conformal invariance in the bulk is broken by the 
(s)quark vacuum expectation values proportional to
$ \sqrt{\xi}$ (the Fayet-Iliopoulos term).}. Thus, our conjecture above is equivalent
to the assumption that the mass of quarks and gauge bosons $m$ 
has singularities  as a function of $g^2$  at $g^2 \sim 1$  in the bulk theory.

The global symmetry of the world-sheet theory we obtained is
\beq
{\rm SO}(3,1)\times {\rm SU}(2)\times {\rm SU}(2)\times {\rm U}(1),
\label{globgroup}
\eeq
where the first factor is the Lorentz group, while the other factors represent global internal symmetries
of the weighted $CP(2,2)$ model ($N=\tN=2$).

Our terminology ``thin string'' should be understood with care. 
The target space of the weighted  $CP(2,2)$ model is a {\em non-compact} Calabi-Yau manifold. 
Since the non-compact
string moduli $\rho^K$ have the string-size interpretation one might think that at large $|\rho |$ our string is not thin. What we mean by the thin-string condition 
(\ref{thinstring}) is that the string core  is thin, and higher-derivative corrections run 
in powers of  ${\pt^2}/{m^2}$ and are negligible.

Note that  there are massless states in the bulk theory ($\tilde{N}$ quarks out of the total set of $N_f$ quarks) which give rise to the continuous spectrum.
Most of  these light modes are {\em not} localized on the string. The only zero modes which are localized
 (in addition to the  translational modes) are the size and the orientational modes \cite{SYsem} indicated in
(\ref{wcp}). All other localized modes are massive 
with mass $\sim m$. Integrating out these massive modes leads
to higher-derivative corrections running in powers of  ${\pt^2}/{m^2}$. They are negligible if   
$m$ is large, see (\ref{thinstring}). We do {\em not} integrate out zero modes.

{\em Gauge-string duality.}---Abstracting ourselves from the solitonic origin of our
model (\ref{wcp}) and (\ref{stringaction}) we can view it as a perfectly legitimate critical string theory,
with all ensuing consequences. 
To describe physics at weak coupling we use the bulk theory in its original formulation in terms of quarks 
and gauge bosons.  As was already mentioned, the quarks and gauge bosons  have 
masses of the order of $m\sim g\sqrt{T}$. At   $g^2\ll 1$ they are light, while stringy states 
with masses of the order of $\sqrt{T}$ are heavy. 

Since it is impossible at the moment to determine the behavior of $m$ versus the bulk coupling $g$ in the 
strong coupling regime it seems reasonable to 
conjecture that at strong coupling  the condition (\ref{thinstring}) is satisfied --
quarks and gauge bosons are heavy while the stringy states are light.  To describe physics in the regime of large
$m$ we should use string theory (\ref{stringaction}). We call this 
gauge-string duality.

Strings in the $U(N)$ theories are stable; they cannot be broken. Thus, we  deal with
the closed string. A detailed study of the closed string spectrum in our string theory (\ref{stringaction})
is left for the future work. Here we limit ourselves to a short comment.

\begin{figure}
\begin{center}
\includegraphics[scale=0.5]{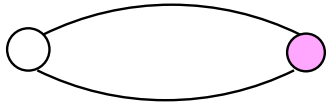}
\end{center}
\caption{\label{figmeson}\small Monopole-antimonopole  stringy meson.
Open and closed circles denote the monopole and antimonopole, respectively.}
\end{figure}

The adjoint (in the $SU(2)$ factors  in (\ref{globgroup})) states are formed by 
monopole-antimonopole pairs
connected by two confining non-Abelian strings (Fig.~\ref{figmeson}),  see  \cite{1} for 
details\,\footnote{This picture assumes
an infrared regularization such that the vacua of the world-sheet theory are discrete.}. 
These  mesons   play an important role in the so-called ``instead-of-confinement'' phase (see
recent review  \cite{SYdualrev}) and at strong coupling should be described as states of a closed string of the string theory 
(\ref{stringaction}). In particular, we expect them to lie on the linear Regge trajectories even at small spins.

These mesons  are quite similar to mesons
in   real-world QCD: they have  the ``correct'' (adjoint or singlet)
 quantum numbers with respect to the unbroken global
group. Moreover,
at strong coupling the monopole-antimonopole meson
becomes stable because it is light and cannot decay into
 screened quarks and gauge bosons.   

Summarizing, we identified four-dimensional Yang-Mills theory which supports a topological BPS soliton --
the string in four dimensions which can be treated as a critical ten-dimensional superstring. The world-sheet theory on the string is described by $R^4\times WCP(2,2)$. The moduli of the string under consideration resulting in 10D critical string construction are as follows: four translational moduli associated with the flat $R^4$
space, plus  six Êorientational moduli associated with the $WCP(2,2)$ target
space. 
In a physical gauge, say the light-cone gauge, two translational modes are
``gauged away" and we are left with two transverse translational modes. The other two translational modes can be readily ``gauged in."

\vspace{2mm}

\section*{Acknowledgments}

We are grateful to Zohar Komargodski for very useful communications and insights.

The work of M.S. is supported in part by DOE grant DE-SC0011842. 
The work of A.Y. was  supported by William I. Fine Theoretical Physics Institute,   
University of Minnesota,
by Russian Foundation for Basic Research Grant No. 13-02-00042a and by Russian State Grant for
Scientific Schools RSGSS-657512010.2. The work of A.Y. was supported by the Russian Scientific Foundation 
Grant No. 14-22-00281.

\end{document}